\documentclass[prd,fleqn,useAMS,twocolumn,superscriptaddress,nofootinbib]{revtex4-1}

\usepackage{graphicx}
\usepackage{amsmath}
\usepackage{amssymb}

\usepackage{hyperref}
\usepackage{tikz}

\hypersetup{hidelinks=True}
\usepackage{natbib}
\usepackage[normalem]{ulem}

\newcommand\numberthis{\addtocounter{equation}{1}\tag{\theequation}}

\newcommand*\diff{\mathop{}\!\mathrm{d}}
\newcommand*\diffcubed{\mathop{}\!\mathrm{d}^3}

\newcommand\numbers[1]{\textcolor{black}{$#1$}}

\newcommand\cov{\mathcal{C}}

\DeclareGraphicsExtensions{.png,.jpg,.pdf}
\begin{document}
\title[2D{$\times$}3D]{Cross-correlating 2D and 3D galaxy surveys}

\author{Samuel Passaglia}
\email{passaglia@uchicago.edu}
\affiliation{Kavli Institute for Cosmological Physics, and the Department of Astronomy \& Astrophysics, University of Chicago, Chicago, IL 60637}
\author{Alessandro Manzotti}
\affiliation{Kavli Institute for Cosmological Physics, and the Department of Astronomy \& Astrophysics, University of Chicago, Chicago, IL 60637}
\author{Scott Dodelson}
\affiliation{Fermi National Accelerator Laboratory, Batavia, IL 60510-0500}
\affiliation{Kavli Institute for Cosmological Physics, and the Department of Astronomy \& Astrophysics, University of Chicago, Chicago, IL 60637}

\date{\today}

\label{firstpage}

\begin{abstract}

Galaxy surveys probe both structure formation and the expansion rate, making them promising avenues for understanding the dark universe. Photometric surveys accurately map the 2D distribution of galaxy positions and shapes in a given redshift range, while spectroscopic surveys provide sparser 3D maps of the galaxy distribution. We present a way to analyse overlapping 2D and 3D maps jointly and without loss of information. We represent 3D maps using spherical Fourier-Bessel (sFB) modes, which preserve radial coverage while accounting for the spherical sky geometry, and we decompose 2D maps in a spherical harmonic basis. In these bases, a simple expression exists for the cross-correlation of the two fields. One very powerful application is the ability to simultaneously constrain the redshift distribution of the photometric sample, the sample biases, and cosmological parameters. We use our framework to show that combined analysis of DESI and LSST can improve cosmological constraints by factors of ${\sim}1.2$ to ${\sim}1.8$ on the region where they overlap relative to identically sized disjoint regions. We also show that in the overlap of DES and SDSS-III in Stripe 82, cross-correlating improves photo-$z$ parameter constraints by factors of ${\sim}2$ to ${\sim}12$ over internal photo-$z$ reconstructions.

\end{abstract}

\maketitle

\section{Introduction}
	\label{sec:introduction}
	The $\Lambda$CDM model is successful in explaining nearly all cosmological observations, yet we lack an underlying physical theory that produces this phenomenological model. The vacuum energy density $\Lambda$ is subject to the well-known cosmological constant problem \citep{Weinberg1989}, and cold dark matter (CDM) could arise from a multitude of extensions to the standard model of particle physics \citep{Feng2010}.

	Galaxy surveys will shed light on whether the cosmological constant is indeed responsible for the current epoch of acceleration. Galaxy clustering and gravitational lensing are sensitive both to geometric expansion and to the growth rate of cosmic structure, and models for cosmic acceleration make a wide variety of predictions for expansion and structure growth. As a result, galaxy surveys will test $\Lambda$CDM and its rivals with unprecedented precision over the coming decade.

	Galaxy surveys can be broadly classified as either spectroscopic or photometric. Spectroscopic surveys measure individual spectra for the galaxies they observe and therefore provide three-dimensional galaxy density maps. Photometric surveys image galaxies in a handful of color bands, often with high resolution, and therefore measure the angular positions and shapes of many more galaxies, with the caveat that the radial position of each is only estimated based on its colors. The resulting photometric redshifts are typically accurate at the 1-10 percent level.

	Therefore, rather than treat each galaxy individually, analysts of photometric surveys typically divide objects into several photometric redshift bins for analysis, treating each bin as a 2D field \citep{Chang2016}. The properties of these bins still depend on the photometric redshift properties of the survey, and therefore their constraining power is often limited by the accuracy of the photometric redshifts. 

	Some patches of the sky are observed by both photometric and spectroscopic surveys. For example, the ongoing Dark Energy Survey \citep{TheDarkEnergySurveyCollaboration2005} and the completed Sloan Digital Sky Survey-III \citep{Alam2015} overlap in a portion of the sky known as Stripe 82, while surveys from the upcoming Large Synoptic Survey Telescope \citep{LSSTScienceCollaboration2009} and Dark Energy Spectroscopic Instrument \citep{DESICollaboration2016} will overlap over roughly ${\sim} 3000 \ \text{deg}^2$ of equatorial sky.

	In these regions, a combined analysis of the overlapping photometric and spectroscopic surveys can help overcome the limitations of both types of surveys. In particular, because overlapping surveys sample overlapping matter distributions, we expect a cross-correlation of the two samples to be sensitive to the redshift distributions of the photometric survey's redshift bins, the galaxy biases of the samples, and cosmological parameters. 
	
	This raises a question: Galaxy clustering in spectroscopic surveys is typically analyzed in three dimensions, while photometric probes, such as the lensing field, are best described in two dimensions. What is the best way to obtain information from a 2D probe that overlaps with a 3D probe? If 2D fields are described by the angular correlation function and spectroscopic fields by the 3D correlation function, what is the relevant statistic describing the cross-correlation between them?

	Indeed, several groups (see, e.g., \citet{Newman2008,McQuinn2013,Menard2013}) have already demonstrated that the cross-correlation between the 2D photometric surveys and 3D spectroscopic data is a powerful way of understanding the errors on photometric redshifts.
	They typically exploit the cross-correlation by working in several steps. A cosmology is first assumed or measured from the surveys individually. This cosmology is then used to compute cross-correlations of the 2D photometric density bins with the 3D spectroscopic galaxy positions, which they also bin into many 2D redshift slices. These cross-correlations are then used to constrain the photo-$z$ and bias parameters of the 2D survey. Finally, these nuisance parameter constraints improve the cosmological constraints from the 2D shear measurements.

	Other groups (see, e.g., \citet{Gaztanaga2012,Kirk2013,Eriksen2015b}) approach the problem using a one-step analysis of the overlapping photometric and spectroscopic surveys, simultaneously constraining cosmology, photo-$z$s, and biases. Again, they treat the 3D survey as a set of 2D redshift slices, which potentially discards some of the radial modes in the 3D survey. \cite{DePutter2013} and \cite{Font-Ribera2014} try to recover these 3D modes by including a small-scale power spectrum $P(k)$.

	In this paper, we present a framework for conducting a one-step analysis of overlapping 2D and 3D surveys while retaining all radial information. As an application of this formalism, we quantify the cosmological benefit of having LSST and DESI overlap as well as possible photo-$z$ constraints from the overlap of DES and SDSS. However, we argue that the formalism itself may prove useful in a wide variety of cosmological analyses as galaxy surveys and cosmic microwave background experiments, each of which have a multitude of probes of the underlying matter distribution, become more powerful.

	Our formalism is based on decomposing 2D fields into spherical harmonics and  3D fields into spherical Fourier-Bessel (sFB) modes. The sFB decomposition was pioneered in cosmology by \cite{Fisher1994}, \cite{Zaroubi1995}, and \cite{Heavens1995} and has seen renewed attention in works by \cite{Shapiro2011}, \cite{Dai2012}, \cite{Rassat2012}, and \cite{Pratten2014}, among others. This one-step approach carries advantages: it does not require a fixed cosmology in the initial determination of photo-$z$ properties; it counts the modes that constrain both photometric properties and cosmological parameters once and only once, where two-step approaches run the risk of counting modes twice (if the same modes are used separately to constrain photometric redshifts and cosmological parameters) or not at all (if the covariance between the 2D and 3D fields is not used in the cosmological analysis).

	We develop our framework in Sec.~\ref{sec:formalism}. We then apply this formalism in a Fisher matrix forecast in Sec.~\ref{sec:fisher}, results of which for LSST-DESI and DES-SDSS are presented in Sec.~\ref{sec:results}. Finally, we conclude in Sec.~\ref{sec:conclusion}.

\section{The 2D $\times$ 3D Formalism}
	\label{sec:formalism}

	\subsection{Decompositions and correlations of 2D and 3D fields}
		\label{subsec:basics}
		We wish to extract all the information present in overlapping 2D and 3D scalar fields. In this work, the 2D fields will be tomographically binned density or shear fields from a photometric survey, while the 3D field will be a density field from a spectroscopic survey.

		Let $A(\hat{n})$ be an observed 2D scalar field, the sky-projection of the true matter density contrast $\delta(r \hat{n})$ weighted by some efficiency kernel $F_A(r)$:
		\begin{equation}
			\label{eq:2dprojection}
			A(\hat{n}) = \int \diff{r} F_A (r) \delta(r \hat{n}),
		\end{equation}
		where $r$ is the line-of-sight comoving distance, and we assume spatial flatness throughout. We will consider two kinds of scalar fields, the galaxy density field with $A=g$ and the lensing convergence field with $A=\kappa$. The efficiency kernel for galaxy clustering is
		\begin{equation}
		F_{g}(r) \equiv b(r) \frac{H(z)}{c} \frac{d n}{d z}\bigg\rvert_{z=r},
		\end{equation}
		while the efficiency kernel for weak lensing is
		\begin{equation}
		F_{\kappa}(r) \equiv  \frac{3}{2} \frac{H_0^2}{c^2} \Omega_m^0 \frac{r}{a(r)} \int_{z(r)}^\infty \diff{z_s}  \frac{d n}{d z_s}  \big[ 1-\frac{r}{r(z_s)}\big].
		\end{equation}		
		$b(r)$ is the scale-independent linear bias evaluated at the time when a photon that reaches us today was a comoving distance $r$ away from us. $\frac{d n}{d z}$ is the field's redshift-space selection function, normalized such that $\int \diff{z} \frac{dn}{dz} = 1$.

		We will absorb all partial sky coverage into a factor $f_{\text{sky}}$ in the Fisher matrix. We refer the reader to \cite{Lanusse2015} for a more detailed discussion of the mode-mixing that occurs with proper treatment of angular masks.
		
		The 2D field $A(\hat{n})$ lives on the sky, and therefore we can expand it in a set of complete basis functions on the sphere, the orthonormal spherical harmonics $Y_{\ell m} (\hat{n})$. We then have:
		\begin{equation}
			\label{eq:2dexpansion}
			A(\hat{n}) = \sum_{\ell,m} A_{\ell m} Y_{\ell m} (\hat{n}).
		\end{equation}
		The spherical harmonic coefficients can be directly computed from the data (see, e.g., \cite{Gorski2005}). We relate these observed quantities to the underlying density field by multiplying each side of Eq.~(\ref{eq:2dexpansion}) by $Y_{\ell'm'}^*(\hat n)$ and integrating over the solid angle $\diff{\Omega}$ corresponding to $\hat n$ :
		\begin{align*}
			\label{eq:2dcoef}
			A_{\ell m} = {} & \int \diff{r} F_{A}(r) \int \diff{\Omega} Y^*_{\ell m} (\hat{n}) \delta(r \hat{n}) \\
			= {} & \int \diff{r} \int \frac{\diffcubed \vec{k}}{(2 \pi)^3} \tilde\delta(\vec{k}) 4 \pi i^\ell j_\ell (kr) Y^*_{\ell m}(\hat{k}) F_{A}(r,k).
			\numberthis
		\end{align*} 
		To get the second equality, we used the spherical harmonic form of the plane wave expansion,
		
		\begin{equation}
			\label{eq:planewave}
			e^{i \vec{k} \cdot \vec{x}} = 4 \pi \sum_{\ell, m} i^{\ell} j_\ell (k x) Y^*_{\ell m} (\hat{k}) Y_{\ell m} (\hat{x}),
		\end{equation}
		and we defined $F_A(r,k) \equiv D(r,k) F_A(r)$, where $D(r,k)$ is the scale-dependent growth function, in order to identify $\tilde\delta(\vec k)$ with a Fourier mode of the present-day matter density constrast.

		Cosmological information is present in the correlations of this field. Since $\langle \tilde\delta(\vec k) \tilde\delta(\vec k')\rangle =  (2\pi)^3 \delta^D(\vec k + \vec k') P(k)$, where $P(k)$ is the matter power spectrum today, the ensemble average of the expansion coefficients of two 2D fields $A(\hat{n})$ and $B(\hat{n})$ is
		\begin{equation}
		\langle A_{\ell m} B^{*}_{\ell'm'} \rangle =  \delta_{\ell\ell'} \delta_{mm'} C_\ell^{A B},
		\end{equation}
		with the angular spectrum $C_\ell^{A B}$ equal to:
		\begin{align*}
			\label{eq:2d2d}
			C_\ell^{A B}= {} & \Bigg[ \frac{2}{\pi}\,\int \diff{r}_1 \int \diff{r}_2 \int k^2 \diff{k} P(k) 
			\\   \times & F_{A}(r_1,k) F_{B}(r_2,k) j_\ell (kr_1) j_\ell(kr_2)+ N_\ell^{A} \delta_{A B} \Bigg].\numberthis
		\end{align*}
		We have added shot noise in the case where $A$ and $B$ are the same field. With $n_{\text{2D}}$ the number of galaxies per steradian in the sample, the noise for galaxy clustering is $N^{g}_\ell =1/n_{\text{2D}}$ and the noise for weak lensing is $N^{\kappa}_\ell = .3^2/(2 n_{\text{2D}})$ (e.g., see \citet{Hearin2012}).

		We can follow \cite{Loverde2008} and do the above radial integrals in the Limber approximation at the expense of accuracy at small $\ell$ by making the substitution $j_\ell (kr) \to \sqrt{\frac{\pi}{2 \ell+1}} \delta^D(kr - \ell-1/2)$. Thus:
		\begin{align*}
			\label{eq:2d2dLimber}
			C_\ell^{A B} \simeq {} & \int \diff{k} \frac{P(k)}{\ell+\frac{1}{2}} F_{A}(\frac{\ell+\frac{1}{2}}{k},k) F_{B}(\frac{\ell+\frac{1}{2}}{k},k) \\ + & N_\ell^{A} \delta_{AB}.\numberthis
		\end{align*}
		A photometric survey will in general have several 2D tomographic bins, and every bin can have galaxy-galaxy, lensing-lensing, and galaxy-lensing information. There will also be information in the correlations across bins. All of these correlations can be computed with the appropriate kernels in Eq.~(\ref{eq:2d2dLimber}). For Gaussian fields, these two-point functions contain all of the cosmological information in the 2D fields.

		Now consider a 3D galaxy over-density field $\beta(\vec{x})$ that overlaps with these 2D fields. Just as we expanded 2D fields in a basis on the sphere, we wish to expand the 3D field in a family of 3D basis functions for scalar fields.

		We anticipate cross-correlating our expansion of the 3D field with the spherical harmonic expansions of the 2D fields, and thus we retain spherical harmonics as the angular basis functions. Motivated by the plane wave expansion, Eq.~(\ref{eq:planewave}), we include the spherical Bessel functions as the radial basis functions. Any cosmological field should be well behaved at the origin, so we need use only the spherical Bessel functions of the first kind, $j_\ell (k r)$. Together, the spherical Bessel functions and spherical harmonics form the spherical Fourier-Bessel (sFB) basis. This basis is the solution set to the Helmholtz equation in spherical coordinates with a nonsingular boundary condition at the origin. 

		The sFB basis has been used in the context of galaxy surveys before. \cite{Leistedt2012} provide a code to quickly compute sFB decompositions from data, and \cite{Pratten2014} study the cross-correlation of the 2D thermal SZ map with the 3D weak lensing potential in a similar language. \cite{Lanusse2015} conduct a detailed study of the benefits of analysing 3D galaxy surveys in the sFB basis. Here we apply this formalism to overlapping 2D and 3D galaxy surveys.

		Consider the 3D galaxy density field:
		\begin{equation}
		\beta(r \hat n) = \phi_\beta(r) b_\beta(r) \delta(r \hat n),
		\end{equation}
		where $\phi_\beta(r)$ is the comoving-space survey selection function, normalized so that $\int \diffcubed{\vec{r}} \phi_\beta(r) = V_\beta$, the volume of the survey in comoving units, and again neglecting partial sky coverage. 
		Expanding in the sFB basis, we have:
		\begin{equation}
			\label{eq:3d density field}
			\beta(r \hat{n}) = \sqrt{\frac{2}{\pi}}  \sum_{\ell,m} \int_0^\infty q \diff{q} \ \beta_{\ell m} (q) Y_{\ell m} (\hat{n}) j_\ell (q r).
		\end{equation}
		The $j_\ell$'s satisfy the following orthogonality condition:
		\begin{equation}
			\label{eq:jl orthonormal}
			\int_0^\infty r^2 \diff{r} j_\ell(q_1 r) j_\ell(q_2 r) = \frac{\pi}{2 q_1^2} \delta^D (q_1 - q_2),
		\end{equation}
		which allows us to solve for the expansion coefficients
		\begin{align*}
			\label{eq:sfb coefficients}
			\beta_{\ell m}(q) = {} & \sqrt{\frac{2}{\pi}} \ \bigg[ \int r^2 \diff{r} \int \diff{\Omega} \\ & \times q \ Y^*_{\ell m} (\hat{n}) j_\ell (q r) \phi_\beta(r) b_\beta(r) \delta(r\hat{n}) \bigg].
			\numberthis
		\end{align*}
		We then express the coefficient in terms of the Fourier transform of the field and introduce the growth function to obtain:
		\begin{equation}
			\label{eq:3dcoef}
			\beta_{\ell m}(q) = \sqrt{\frac{2}{\pi}} \int \frac{ \diffcubed  \vec{k}}{(2 \pi)^3} \tilde\delta(\vec{k}) 4 \pi i^\ell Y^*_{\ell m} (\hat{k}) W^\beta_{\ell}(k,q),
		\end{equation}
		where we have defined
		\begin{equation}
			\label{eq:W}
			W^\beta_{\ell}(k,q) \equiv q \int \tilde{r}^2 \diff{\tilde{r}} j_\ell (q\tilde{r}) j_\ell (kr) D(r,k) b_\beta(r) \phi_\beta(r).
		\end{equation}
		The $\tilde{r} \equiv r_{\text{fid}}(z)$ that appears in Eq.~(\ref{eq:W}) is the distance computed using the fiducial redshift-distance relation, which appears because spectroscopic surveys measure redshifts rather than distances. See Appendix~\ref{app: fiducial} for a derivation and discussion of this subtlety.

		Cosmological information is in the autocorrelation of the field: 
		\begin{equation}
		\langle \beta_{\ell m}(q) \beta^*_{\ell'm'}(q') \rangle = \delta_{\ell\ell'} \delta_{mm'}	C^{\beta\beta}_\ell(q,q'), 
		\end{equation}
		with the autospectrum
		\begin{align*}
			\label{eq:3d3d}
			C^{\beta\beta}_\ell(q,q') 
			\equiv {} & \left(\frac{2}{\pi}\right)^2 \int k^2 \diff{k} P(k) W^\beta_{\ell}(k,q) W^\beta_{\ell}(k,q') \\ 
			& + \frac{2}{\pi} \frac{q q' }{n_{3D}} \int \phi_\beta(r) j_\ell(q r) j_\ell (q' r) r^2 \diff{r}.
			\numberthis
		\end{align*}
		$n_{\text{3D}}$ is the survey's mean number of objects per comoving megaparsec, $\frac{N}{V}$. The galaxy shot noise term is derived in Appendix C of \cite{Lanusse2015} as well as in \cite{Yoo2013}, who do a fully relativistic treatment of galaxy clustering in the sFB framework. For an isotropic and homogeneous Gaussian field, $C^{\beta\beta}_\ell(q,q')$ contains all of the cosmological information in the 3D clustering field. 

		One advantage of this formalism is its ability to handle the cross-correlation of the 2D and 3D fields while retaining all of the 3D field's radial information. In particular,
		\begin{equation}
		\langle A_{\ell m} \beta^*_{\ell'm'}(q) \rangle = C_\ell^{A \beta}(q) \delta_{\ell\ell'} \delta_{mm'}
		\end{equation}
		with the cross-spectrum
		\begin{equation}
		C_\ell^{A\beta}(q) \equiv \left(\frac{2}{\pi}\right)^{\frac{3}{2}} \int k^2 \diff{k} P(k) W^\beta_{\ell}(k,q) \int \diff{r} F_A(r,k) j_\ell (kr).
		\end{equation}
		We can again compute the radial integral in the Limber approximation, leaving a very simple expression for the cross-correlation:
		\begin{equation}
		\label{eq:2d3dLimber}
		C_\ell^{A\beta}(q) \simeq  \frac{2}{\pi} \frac{1}{\sqrt{\ell}}  \int k \diff{k} P(k) W^\beta_{\ell} (k,q) F_A(\frac{\ell}{k},k).
		\end{equation}
		We now have a formalism in which to jointly analyze overlapping 2D and 3D samples without loss of information. All the information in the overlapping samples is contained in the 2D correlations [Eq.~(\ref{eq:2d2dLimber})],  the 3D auto-correlation [Eq.~(\ref{eq:3d3d})], and the 2D-3D cross-correlations [Eq.~(\ref{eq:2d3dLimber})].

	\subsection{Discretization of the radial degree of freedom}
		\label{subsec:discretization}

		Redshift surveys have limited radial extent, and therefore we can impose the boundary condition that $\phi(r \geq r_{\text{max}}) = 0$. In this work, the sharp redshift cutoffs of our 3D samples set $r_{\text{max}}$. This boundary condition provides us with a useful discretization of the spherical Bessel function's radial degree of freedom $q$, because only spherical Bessel functions with zeros at $r_{\text{max}}$ will be involved in the expansion. For a given $\ell$, the previously continuous $q$ spectrum is now discretized at $n_\text{max}(\ell)$ different values $q_{\ell n} = \frac{\zeta_{\ell n}}{r_{\text{max}}}$, where $\zeta_{\ell n}$ denotes the $n$'th zero of the $\ell$'th spherical Bessel function.

		This limited radial extent breaks the normalization of the $j_\ell$'s. Rather than the normalization in Eq.~(\ref{eq:jl orthonormal}), we now have from \cite{Fisher1994} that:

		\begin{equation}
		\int_0^{r_{\text{max}}} r^2  \diff{r} j_\ell (q_{\ell n} r) j_\ell (q_{\ell  n'} r) = \delta_{nn'} \frac{r_{\text{max}}^3}{2} \left[j_{\ell+1} (q_{\ell n} r_{\text{max}})\right]^2.
 		\end{equation}
 		Because of this different normalization, in order to preserve the form of the coefficients in Eq.~(\ref{eq:sfb coefficients}) we must write the transform of the density field, Eq.~(\ref{eq:3d density field}), as:
		\begin{equation}
			\beta(r \hat{n}) = \sum_{\ell,m,n} \frac{\sqrt{2 \pi} q_{\ell n} \beta_{\ell m} (q_{\ell n}) Y_{\ell m} (\hat{n}) j_\ell (q_{\ell n} r)}{r_{\text{max}}^3 [j_{\ell+1} (q_{\ell n} r_{\text{max}})]^2 }.
		\end{equation}
		With Eq.~(\ref{eq:sfb coefficients}) for the coefficients preserved, all the results in the previous section still hold, and the spectra $C_\ell^{A\beta}(q_{\ell n})$ and $C_\ell^{\beta\beta}(q_{\ell n}, q_{\ell n'})$ are now discrete.

	\subsection{Representations of physical scales}
		\label{subsec:linear}

		Shell-crossing and baryonic effects complicate galaxy clustering at small physical scales, the nonlinear regime. It is therefore important to understand how a given physical scale $k$ is probed by the decompositions we have presented.

		When decomposing a 2D field, Eq.~(\ref{eq:2dcoef}), the radial integral gets its largest contributions from $k r \simeq \ell$ for large $\ell$. Therefore, in a narrow tomographic bin with a median redshift $z_{\text{med}}$, a physical scale $k$ is probed by the set of $A_{\ell m}$ with $\ell \simeq k r(z_{\text{med}})$.

		\begin{figure}
		\includegraphics[width=.45\textwidth]{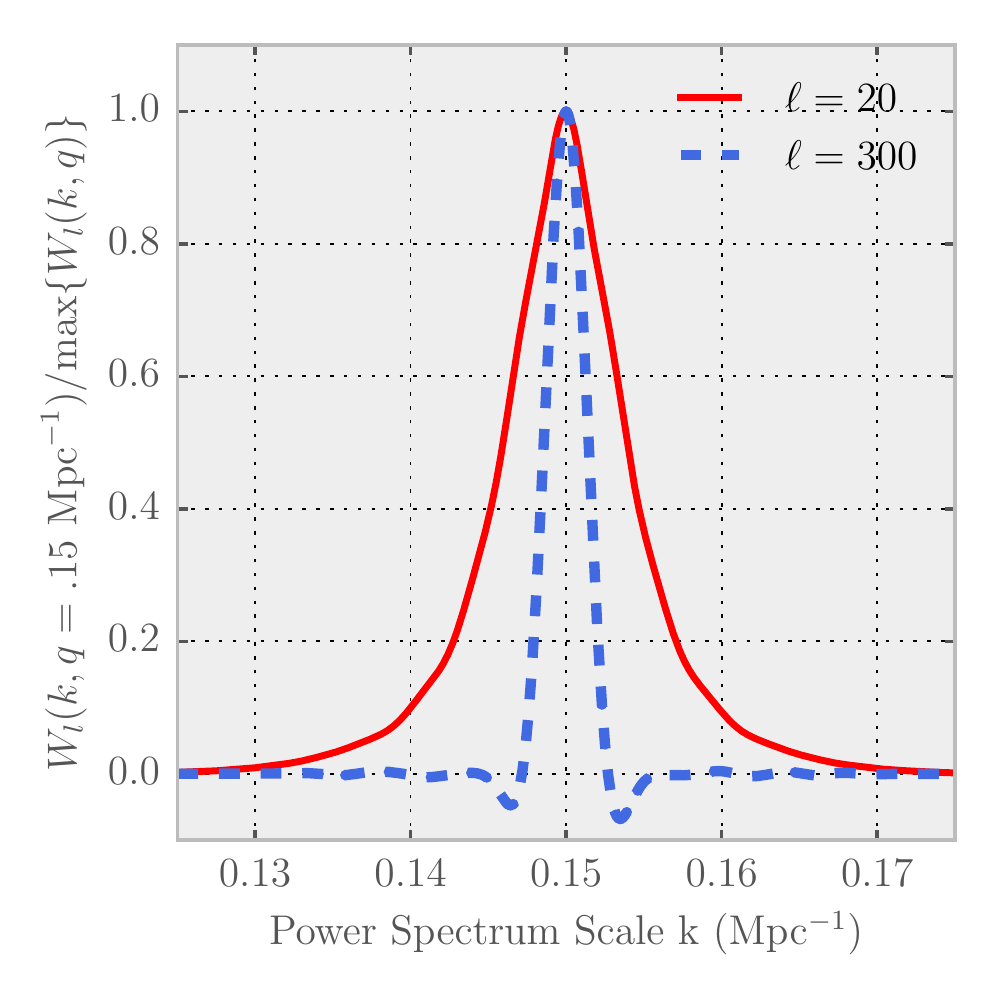}
		\caption{Physical scales $k$ probed by our choice of $q_{max}=.15 \ \text{Mpc}^{-1}$, for SDSS. We see that at both low and high l, our maximum sFB radial degree of freedom $q_{max}$ corresponds closely to the maximum physical scale which we wish to consider. For a discussion of the qualitative behavior of $W_\ell(k,q)$, refer to Sec.~\ref{subsec:linear}.}
		\label{fig:W}
		\end{figure}

		In a 3D decomposition, Eq.~(\ref{eq:3dcoef}), the window function $W^\beta_\ell(k,q)$ in Eq.~(\ref{eq:3dcoef}) is sharply peaked around $k \simeq q$. In fact, in the large survey, unity bias, and no growth limit, $W^\beta_\ell(k,q) \to \frac{\pi}{2k} \delta^D(k-q)$. In this limit, the physical scale $k$ is probed only by the set of $\beta_{\ell m}(q)$ with $q = k$.

		In practice, $W^\beta_\ell(k,q)$ will mix scales due to the survey selection function, redshift dependent bias, and linear growth, but Fig.~\ref{fig:W} shows that for the model surveys we consider, a physical scale $k$ is still in large part probed by the data $\beta_{\ell m}(q\simeq k)$. \

		The qualitative behavior of the window functions $W_\ell(k,q)$ in Fig.~\ref{fig:W} can be understood by inspection of the relevant integrals. For a relatively flat survey $dN/dz$, such as the SDSS $dN/dz$ (see Fig.~\ref{fig:stripe82 dndz}), the 3D selection function $\phi(r)$ which enters into $W_\ell(k,q)$ will go as $1/r^2$. This $1/r^2$ weighting of the integral cancels with the explicit $r^2$ factor, and therefore, neglecting bias and growth effects, $W_\ell(k,q)$ becomes an unweighted integral of $j_\ell (qr)$ and $j_\ell (kr)$. The maximum will occur when $k=q$, and the integral will go to zero once the Bessel functions decohere. This will occur roughly once the first peaks of the Bessel functions no longer overlap. As a function of $r$, the width of the first peak of a Bessel function is roughly $1/k$, independent of $\ell$, and the location of the peak is roughly $\ell/k$. Therefore the two Bessel functions destroy each other, and $W_\ell(k,q)$ goes to zero, once $|k-q| \sim q/\ell$. We can therefore expect higher $\ell$ modes to be more sharply peaked, as seen in Fig.~\ref{fig:W}.

		Nonlinear scales are rich in information about cosmology, photo-$z$, and bias. Observers do measure the power spectrum on these scales. Though the framework we have presented is completely capable of handling the nonlinear scales, the importance of scale-dependent bias makes galaxy clustering difficult to interpret on these scales. When forecasting the effectiveness of our framework, we therefore limit our clustering analysis to the quasilinear scale, which we define as $k_{\text{lin}} = .15 \ \text{Mpc}^{-1}$. Our results are insensitive to this choice. The \cite{Takahashi2012} HaloFit power spectrum at this scale should be $\sim{5\%}$ accurate at the redshifts we consider.

		The maximum scale $k_{\text{lin}}$ translates to a maximum angular mode $\ell_{\text{max}} = k_{\text{lin}} r(z_{\text{med}})$ in our 2D clustering decompositions and a maximum radial degree of freedom  $q_{\ell n_{\text{max}}} = k_{\text{lin}}$ in our 3D clustering decompositions. 

		Lensing is sensitive to the underlying matter distribution rather than biased tracers, and therefore we use shear observables on all scales until we become noise-dominated around $\ell \sim 5000$.

		Finally, we only use scales $\ell > \ell_\text{min}=20$ to avoid errors in the Limber approximation, though our results are unchanged with $\ell_\text{min}=10$.

\section{Forecasting}
	\label{sec:fisher}

	We have presented a framework for combined analysis of overlapping photometric and spectroscopic surveys. We would like to apply our framework to quantify the cosmological and photo-$z$ benefits of survey overlap. We therefore need to quantify the information content of an analysis using our framework. The Fisher information matrix  (see, e.g., \cite{Tegmark1997}) allows us to do this. For a set of observables with covariance matrix $\cov$ and parameter-independent mean, the Fisher matrix is composed of elements:
	\begin{equation}
	\label{eq:fisher}
		F_{\mu \nu} = \frac{1}{2} \text{Tr} (\cov^{-1} \cov_{,\mu} \cov^{-1} \cov_{,\nu}),
	\end{equation}
	where $,\mu$ denotes a derivative with respect to the parameter $\theta_\mu$. and $\mu,\nu$ run over the cosmological, photo-$z$, and bias parameters that describe the data.

	The covariance matrices of our observables are block diagonal in $\ell$ and $m$, and blocks at the same $\ell$ are equal for all $m$. We denote these blocks $\cov_\ell$, simplifying Eq.~(\ref{eq:fisher}) to:
	\begin{equation}
	\label{eq:fishersimple}
    	F_{\mu \nu} = \frac{f_{\text{sky}}}{2} \sum_\ell (2\ell+1) \text{Tr} (\cov_\ell^{-1} \cov_{\ell,\mu} \cov_\ell^{-1} \cov_{\ell,\nu})
    \end{equation}
 	where we have included an $f_{\text{sky}}$ factor to account for the angular mask of the survey.

	For a full 2D$\times$3D analysis with $n_T$ tomographic clustering and shear bins in the 2D sample and an overlapping 3D clustering sample, our data vector is comprised of spherical harmonic coefficients for each 2D field and sFB coefficients for the 3D field: $\left\{ p^i_{\ell m}, \beta_{\ell m}(q_{\ell n})\right\}$, where $p \in \{g, \kappa\}$, $i \in [1, n_T]$, and $n \in [1, n_\text{max}(\ell)]$. Thus the covariance matrix $\cov_\ell$ of the data vector is
	\begin{equation}
	\begin{bmatrix}
	\label{eq:cov}
	C_\ell^{p^i q^j} & C_\ell^{p^i \beta}(q_{\ell n'}) \\
	C_\ell^{q^j \beta}(q_{\ell n})^T & C_\ell^{\beta\beta} (q_{\ell n}, q_{\ell n'})
    \end{bmatrix}
    \end{equation}
	where:
	\begin{itemize}
	\item $C_\ell^{p^i q^j} $ is the $2 \cdot n_T \times 2 \cdot n_T$ matrix of auto- and cross-correlations of clustering and lensing in the 2D tomographic bins, computed with Eq.~(\ref{eq:2d2dLimber}).
	\item $C_\ell^{q^j \beta}(q_{\ell n})$ is the $n_\text{max}(\ell)\times 2 \cdot n_T$ matrix of the cross-correlations of the 2D spherical harmonic coefficients with the 3D field's sFB coefficients, computed with Eq.~(\ref{eq:2d3dLimber}).
	\item $C_\ell^{\beta\beta} (q_{\ell n}, q_{\ell n'})$ is the $n_\text{max}(\ell)\times n_\text{max}(\ell)$ matrix of the auto-correlation of the 3D field, computed with Eq.~(\ref{eq:3d3d}).
	\end{itemize}

	The marginalized Gaussian error on parameter $\theta_\mu$ is given by $\sigma_\mu = \sqrt{(F^{-1})_{\mu \mu}}$.

	\subsection{Fiducial models}

		Our fiducial cosmology consists of a flat Planck 2015 $\Lambda$CDM universe, with $\Omega_m = .316$, $\sigma_8 = .8$, $h=.67$, $\Omega_b=.049$, $n_s= .96$, and $w=-1$ \citep{Ade2016}. We compute the linear matter power spectrum in this model using the CAMB package \citep{Lewis2000} and apply the HaloFit prescription \citep{Smith2003} to the linear power spectrum to get the nonlinear power spectrum. We use the \cite{Takahashi2012} HaloFit version, but obtain similar results with the \cite{Mead2015} version.

		Following \cite{Ma2006}, we model photometric redshift errors as Gaussian at each redshift:
		\begin{equation}
		p(z_\text{ph}|z) = \frac{1}{\sqrt{2 \pi} \sigma_z} \exp \left[ - \frac{(z-z_\text{ph}-z_\text{bias})^2}{2 \sigma_z^2} \right].
		\end{equation}
		We allow the photometric redshift bias $z_\text{bias}(z)$ and scatter $\sigma_z(z)$ to be free functions of redshift. In practice, we represent these functions at discrete redshifts, the median of each of the photometric survey's redshift bins, and linearly interpolate to evaluate the function at arbitrary redshifts. The effect of the photo-$z$ uncertainty is to smear tomographic redshift bins, and a sharp bin in photometric redshift space will be diffuse in the space of true redshifts.

		We model the linear bias for each survey as a scale-independent function $b(z)$, which we again represent at the median of the 2D survey's redshift bins and linearly interpolate between bins. Allowing the bias to vary from bin to bin is important because, as highlighted by, e.g. \cite{DePutter2014}, the effect of the redshift evolution of galaxy bias on angular correlations is degenerate with the effect of photo-$z$ errors.

		We consider two different sets of fiducial surveys, one for which a 2D$\times$3D analysis could be done today, and one based on upcoming surveys.

		\begin{figure}
		\includegraphics[width=.45\textwidth]{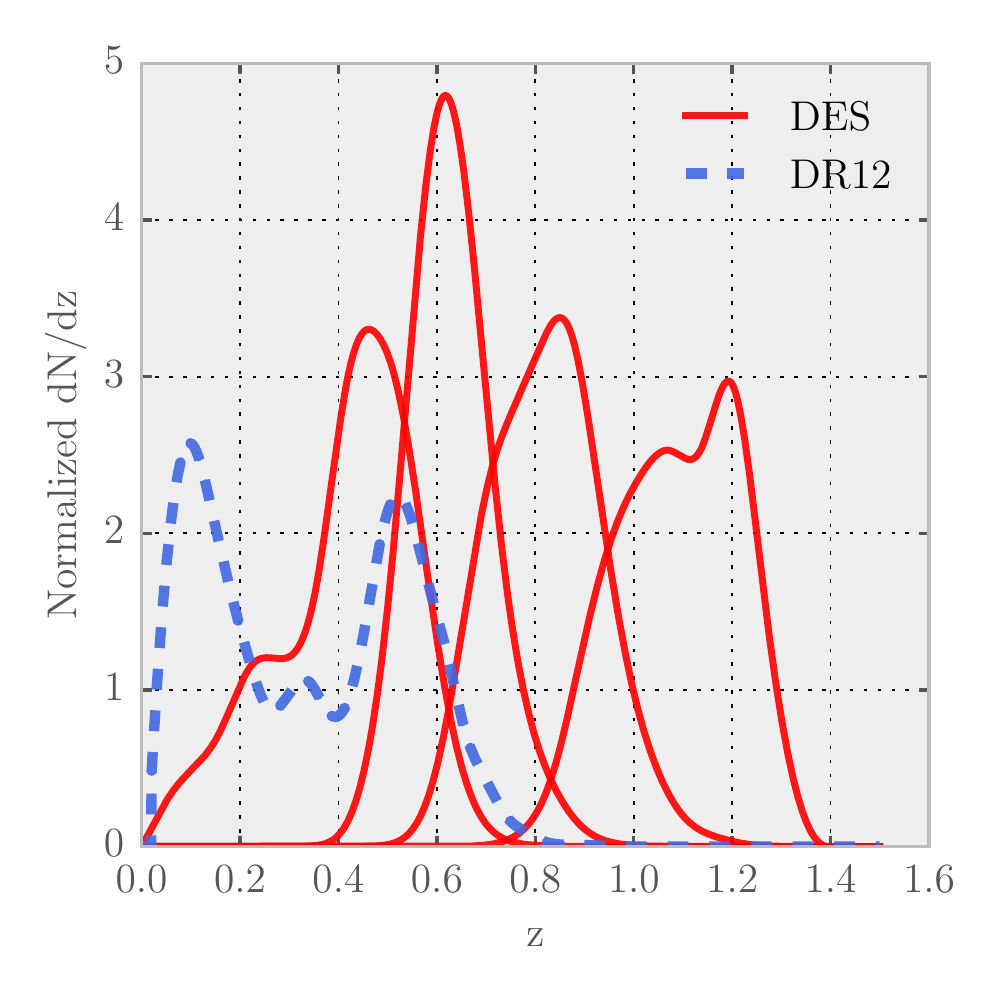}
		\caption{Redshift distributions of the 4 DES tomographic bins (solid red) and the SDSS spectroscopic galaxies (dashed blue). Each tomographic bin has been normalized to have unit integral, as has the 3D distribution.}
		\label{fig:stripe82 dndz}
		\end{figure}

		Our set of contemporary surveys is modeled after the ${\sim}180$ deg$^2$ overlap of the Dark Energy Survey and the Sloan Digital Sky Survey in Stripe 82. We consider only galaxy clustering data for these surveys and are mainly interested in photo-$z$ parameter constraints. The SDSS sample is comprised of SDSS DR12 CMASS and LOWZ spectroscopic galaxies as well as SDSS DR7 galaxies. We treat the SDSS sample using the 3D sFB formalism. The photometric DES data, though not yet publicly available in Stripe 82, will have properties similar to the DES Science Verification data\footnote{http://des.ncsa.illinois.edu/releases/sva1}. We split the DES sample into 4 photo-$z$ bins of equal number density, and we treat each photo-$z$ bin as a 2D sample. The surveys have number density $n^{\text{SDSS}}_\text{2D} = 0.065 \ \text{gal/arcmin}^2$ and $n^{\text{DES}}_\text{2D} = 6 \ \text{gal/arcmin}^2$. In Fig.~\ref{fig:stripe82 dndz}, we plot the normalized redshift distributions of SDSS and DES. At each tomographic bin's median redshift $z_i$, we choose fiducial photo-$z$ parameter values $\sigma_z(z_i) = 0.05 (1+z_i)$ and $z_\text{bias}(z_i) = 0$. We choose fiducial linear bias parameters $b^{\text{DES}} (z_i) = b^{\text{SDSS}} (z_i) = 1/ D(z_i) $ \citep{Bonnett2016, Alam2015}.

		\begin{figure}
		\includegraphics[width=.45\textwidth]{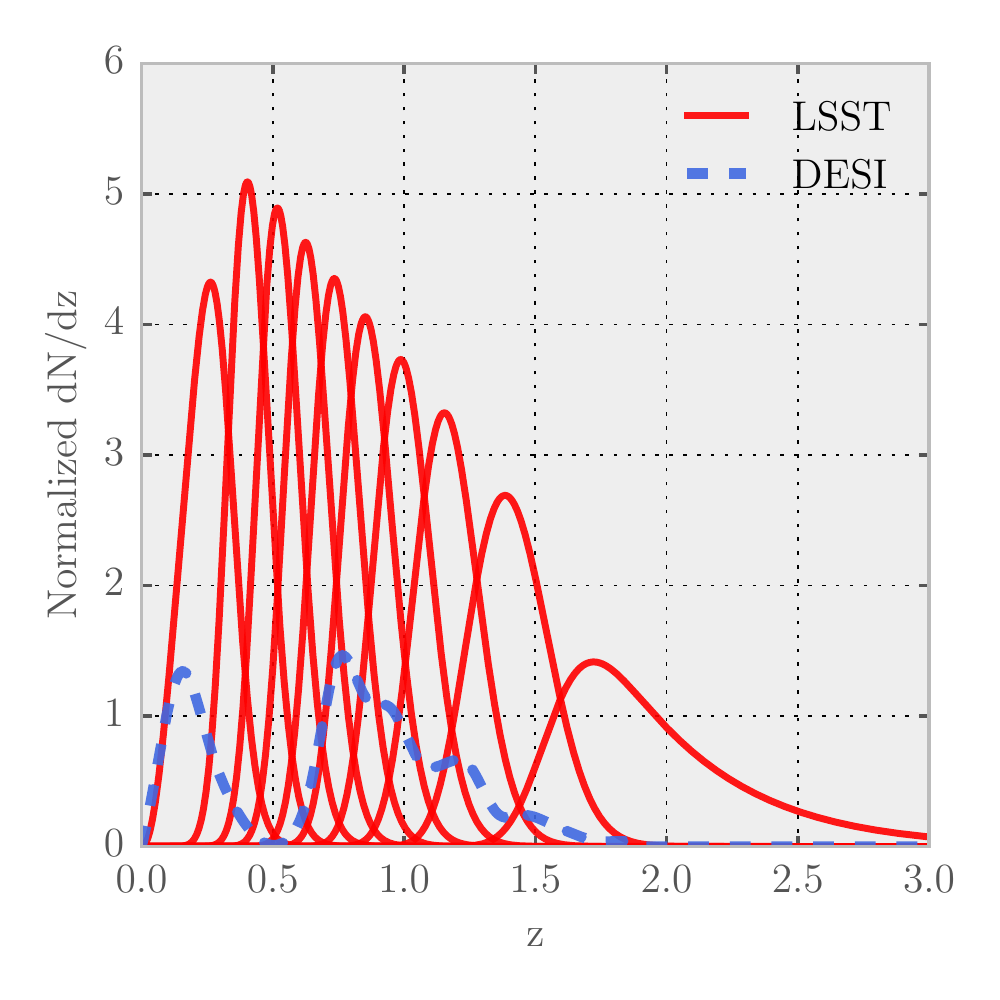}
		\caption{Redshift distributions of the 10 LSST tomographic bins (solid red) and the DESI spectroscopic galaxies (dashed blue). Each 2D tomographic bin has been normalized to have unit integral, as has the 3D distribution.}
		\label{fig:lsst dndz}
		\end{figure}

		Our set of upcoming surveys approximates LSST and DESI. These surveys are expected to overlap in a ${\sim} 3000$ deg$^2$ patch. We split the LSST sample into 10 tomographic bins of equal number density, each with clustering and lensing information. We show the LSST bins and the DESI 3D redshift distribution in Fig.~\ref{fig:lsst dndz}. The LSST sample has $n^{\text{LSST}}_\text{2D} = 50 \ \text{gal/arcmin}^2$, and the DESI sample has $n^{\text{DESI}}_\text{2D} = 0.63 \ \text{gal/arcmin}^2$. We choose fiducial values $\sigma_z(z_i) = 0.05 (1+z_i)$, and  $z_\text{bias}(z_i) = 0.$  We choose fiducial bias parameters $b^{\text{LSST}} (z_i) = b^{\text{DESI}} (z_i) = 1 / D(z_i)$. \citep{LSSTScienceCollaboration2009,DESICollaboration2016,Font-Ribera2014}

		We performed our Fisher matrix analysis on the set of parameters $\Theta = (\Theta_{\text{cosmo}}, \Theta_{\text{bias}},\Theta_{\text{photo-$z$}})$, with $\Theta_{\text{cosmo}}=(\Omega_m, \sigma_8, h, \Omega_b, n_s, w)$  and $\Omega_k=0$ fixed, $\Theta_{\text{bias}} = (b^A(z_i), b^\beta(z_i))$ and $\Theta_{\text{photo-$z$}} = (\frac{\sigma_z(z_i)}{1+z_i}, z_\text{bias}(z_i))$. We do not include any galaxy-shear cross-correlation bias parameters $r_g$ since these should be ${\sim}1$ in the quasilinear regime \citep{Cacciato2012}. We do not impose any external priors.

\section{Results}
	\label{sec:results}

\subsection{Cosmological benefit of LSST/DESI overlap}
		\label{subsec:lsstdesicosmo}
		LSST and DESI will overlap over a large portion of the equatorial sky. We want to understand how this overlap region will improve cosmological parameter constraints. We therefore conduct a combined analysis of these two flagship surveys on the overlap region in the framework we presented in Sec.~\ref{sec:formalism}. We use the Fisher matrix formalism of Sec.~\ref{sec:fisher} to compare parameter constraints from a complete analysis using the covariance matrix from Eq.~(\ref{eq:cov}) to the constraints in which the survey cross-correlation elements are set to zero, i.e. the constraints from disjoint patches.

		\begin{table}[h]
		\bgroup
		\def\arraystretch{1.5}%
		\setlength\tabcolsep{5pt}
		\begin{tabular}{l|c|c|c|c|c|c}
		                                         & $h$   & $\Omega_m$ & $\sigma_8$ & $\Omega_b$ & $w$   & $n_s$ \\ \hline 
		 Full Analysis 							 & $.38$ & $.89$      & $.40$      & $2.3$      & $1.5$ & $.45$ \\
		 LSST Nuisances Fixed                    & $.37$ & $.62$      & $.21$      & $2.2$      & $.83$ & $.40$ \\
		 DESI Nuisances Fixed                    & $.23$ & $.40$      & $.23$      & $2.0$      & $.89$ & $.39$ \\
		 Cosmology Only                          & $.14$ & $.32$      & $.13$      & $2.0$      & $.51$ & $.36$ 

		\end{tabular}
		\egroup
		\caption{Forecasted cosmological constraints ($| \sigma_\theta / \theta | \times 100$) from the $3000$ deg$^2$ overlap of LSST and DESI.}
		\label{tab:lsstdesi}
		\end{table}

		\begin{figure}
		\includegraphics[width=\linewidth]{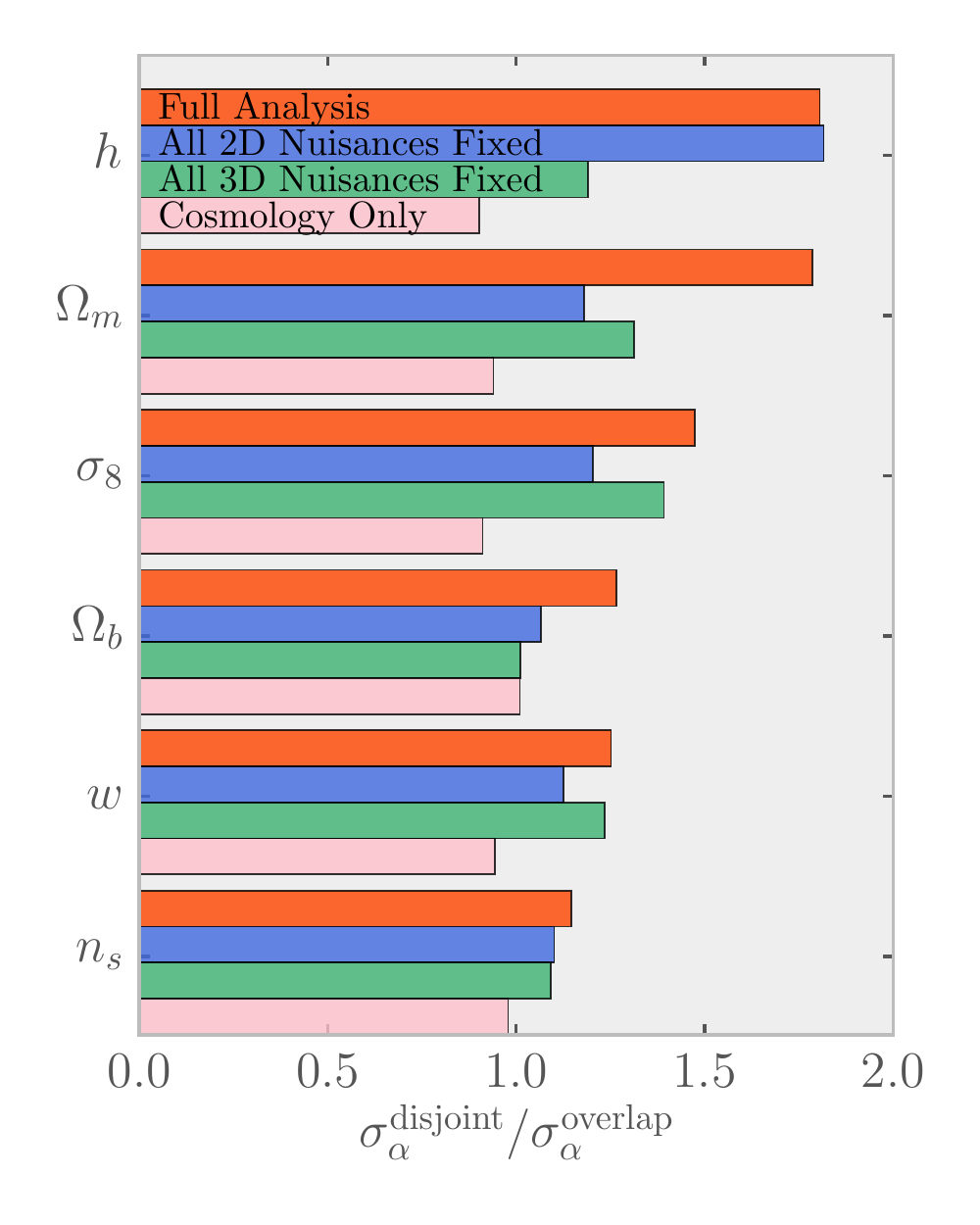}
		\caption{Errors on cosmological parameters for overlapping LSST and DESI patches compared to errors in disjoint patches of the same size. Larger bars indicate a larger same-sky benefit. The red upper bars show the improvements when all parameters are marginalized over, the blue upper middle bars when LSST's nuisance parameters are fixed, the green lower middle bars when DESI's nuisance parameters are fixed, and the pink bottom bars when all nuisance parameters are fixed. When all nuisance parameters are marginalized over, the LSST-DESI overlap will lead to factors of \numbers{{\sim}1.2} to \numbers{{\sim}1.8} improvement in cosmological parameter constraints over disjoint patches. For a discussion of the behavior of the parameters under the fixing of various sets of nuisance parameters, refer to Sec.~\ref{subsec:lsstdesicosmo}.}
		\label{fig:lsst cosmoerrors}
		\end{figure}

		Table \ref{tab:lsstdesi} shows the absolute constraints on cosmological parameters from LSST and DESI in the case of overlapping surveys, while Fig.~\ref{fig:lsst cosmoerrors} shows these constraints relative to the case of disjoint surveys. Our results show that the LSST-DESI overlap will lead to factors of \numbers{{\sim}1.2} to \numbers{{\sim}1.8} improvement in cosmological parameter constraints.

		In order to qualitatively understand the origin of our parameter improvements, Table \ref{tab:lsstdesi} and Fig.~\ref{fig:lsst cosmoerrors} show the absolute constraints and the same-sky improvement factors with various sets of parameters fixed.

		Figure~\ref{fig:lsst cosmoerrors} is perhaps best understood by examining the bars from bottom to top for each cosmological parameter. When all nuisance parameters are fixed, we see that the constraints on cosmological parameters are in fact lower for overlapping surveys than for disjoint surveys by factors of \numbers{{\sim}.9} to \numbers{{\sim}1}. In this case, the cosmological information in the 2D-3D cross-correlation does not fully overcome the double counting of modes that occurs when the surveys overlap.

		Once we include nuisance parameters in our analysis, all cosmological parameters see improvements from survey overlap. The middle lower green bars keep DESI's galaxy biases fixed, but allow LSST's photo-$z$ and bias parameters to vary, while the middle upper blue bars do the reverse. Since nuisance parameters are partly degenerate with cosmological parameters, survey cross-correlation helps to break the degeneracy and improves cosmological parameter constraints.

		By comparing the middle lower green bars with the middle upper blue bars, we can see that, for example, $h$ is much more degenerate with the 3D biases than the 2D photo-$z$s and 2D biases. When the 3D biases are free, in the middle upper blue bar, the benefit of overlapping is dramatic relative to the benefit when the 3D biases are fixed, the green and pink lower bars. In fact, we have checked that much of the constraint on $h$ comes from the autocorrelation of the 3D survey. Since this autocorrelation is sensitive to the 3D survey's 10 bias parameters the survey overlap leads to a significant improvement in $h$ by providing an independent calibration, the 2D-3D cross-correlation. When the 3D bias parameters are fixed, the lower middle green bar, the same-sky benefit is much smaller.

		The top red bars, our final result, show that $h$ and $\Omega_m$ benefit the most from survey overlap. While the improved constraint on $h$ can easily be tied to the improved constraint on the DESI biases that LSST provides, $\Omega_m$ is constrained by both LSST and DESI and thus is sensitive to all nuisance parameters in the problem.

		$\sigma_8$ also shows large improvement with overlap, while $n_s$, $w$, and $\Omega_b$ show more modest overlap improvements. As expected, all the improvements are related to improved photo-$z$ and bias constraints and none of the parameters prefer disjoint surveys.

		Our results are consistent with previous work by \cite{DePutter2013} and \cite{Font-Ribera2014}. They do not include photo-$z$ errors and find only only modest improvement in cosmological constraints from overlapping surveys. The upper middle blue bars in our Fig.~\ref{fig:lsst cosmoerrors} reproduce this result, showing that when we do not marginalize over photo-$z$ errors or 2D biases the same-sky benefit for most parameters is of order \numbers{{\sim}1.1} to \numbers{{\sim}1.2}. The constraint on $h$ shows a much larger same-sky improvement of \numbers{{\sim}1.8}. Since $h$ is primarily constrained by the 3D survey, and they performed their analysis of the 3D sample in redshift slices with an additional small scale $P(k)$ to recover the radial information, we do not expect exact agreement with their work.

		Our result also generally agrees with \cite{DePutter2014}, who showed that if the bias is well behaved, photo-$z$ constraints can significantly improve cosmological constraints from weak lensing. Our work makes the broader statement that even with free bias parameters for each survey in every bin, overlapping surveys lead to important improvements in cosmological constraints.

		We tend to see more modest cosmological parameter improvements than \cite{Kirk2013} and \cite{Jouvel2014}, though direct comparison is not possible as those works represent their spectroscopic survey solely using redshift slices of width $\delta_z = 0.05$, discarding some of DESI's BAO-scale radial information.

		In addition to the direct cosmological parameter improvement on the overlap patch, nuisance parameter constraints derived from a survey overlap region will translate outside the overlap patch and improve cosmological parameter constraints there. 

		In particular, photo-$z$ constraints from the overlapping region will be applicable outside the patch. While we do not include the details of this effect in this work, we can nonetheless examine how photo-$z$ constraints derived from an overlapping survey region compare to internally reconstructed photo-$z$ constraints.

		\begin{figure*}
		\includegraphics[width=\linewidth]{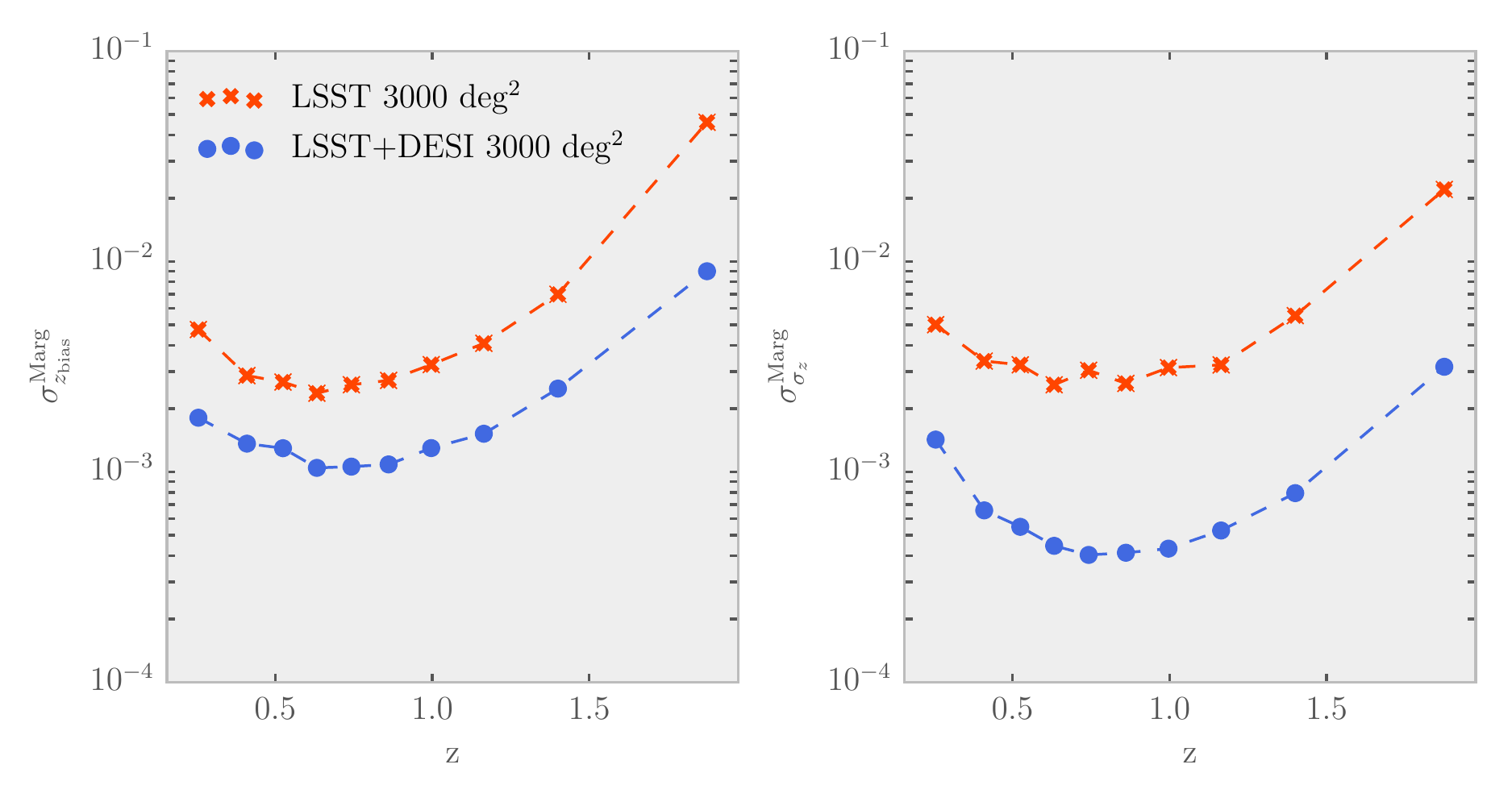}
		\caption{Errors on LSST photo-$z$ parameters on a patch overlapping with DESI (blue circles) and on an isolated patch (red crosses), with cosmological parameters fixed. We connect the discrete parameter constraints with dashed lines as a visual aid. Note the logarithmic scale. Cross-correlations between LSST and DESI on the overlap region shrink marginalized errors on the photo-$z$ parameters by factors ranging from \numbers{{\sim }2} to \numbers{{\sim }8}. Constraints on $\sigma_z$ generally benefit more from the cross-correlation than constraints on $z_\text{bias}$. We fixed all cosmological parameters for this plot, so as not to include any photo-$z$ constraint improvement derived solely from improved cosmological parameter constraints. In practice, marginalizing over cosmological parameters degrades only the internal reconstruction points, not the cross-correlation points.}
		\label{fig:lsst photoz}
		\end{figure*}

		In Fig.~\ref{fig:lsst photoz}, we show that cross-correlating LSST and DESI will produce significant constraints on LSST photo-$z$ parameters, with improvements over constraints from disjoint patches ranging from factors of \numbers{{\sim }2} to \numbers{{\sim }8}. Constraints on $\sigma_z$ generally benefit more from the cross-correlation than constraints on $z_\text{bias}$. We fixed all cosmological parameters for this comparison, so as not to include any photo-$z$ constraint improvement derived solely from improved cosmological parameter constraints. In practice, marginalizing over cosmological parameters degrades only the constraints from LSST internal reconstruction, not constraints including cross-correlation with DESI.

		We note that even where there are few spectroscopic galaxies, photo-$z$ constraints are improved when surveys overlap. In particular, at high $z$, where the 2D-3D overlap is smallest and $z_\text{bias}$ sees its weakest constraint, it also sees its largest improvement relative to the disjoint analysis. With few spectroscopic galaxies overlapping with this high-$z$ 2D bin, the primary way to constrain the photo-$z$ error is by cross-correlating with the adjacent lower redshift 2D bin. This lower-$z$ bin is itself subject to photo-$z$ and bias uncertainties. In the case of overlapping surveys, the lower-$z$ bin overlaps with the 3D survey and therefore the 2D-3D cross-correlation constrains the lower-$z$ bin's photo-$z$ and bias parameters. These improved constraints are then propagated to the high-$z$ bin through the 2D-2D bin cross-correlation.

		This same phenomenon can be seen at $z\sim.5$, where our DESI-like survey has a gap between low-$z$ bright galaxies and high-$z$ emission line galaxies, luminous red galaxies, and quasars. Despite this gap, LSST photo-$z$ errors are well constrained at $z\sim.5$. Though there are plans to fill this gap with an intermediate-$z$ luminous red galaxy sample, our forecast suggests that such plans are not required for LSST photo-$z$ constraints.

		Therefore, using our framework to include all radial information, and account for all photo-$z$ and bias nuisance parameter covariances, we find that there is a benefit to having a DESI-LSST overlap, both in direct cosmological parameter constraints from clustering and weak lensing on the overlap patch and in constraints on superpatch nuisance parameters such as photo-$z$ errors.

	\subsection{Photo-$z$ constraints from DES/SDSS overlap}

		DES and SDSS overlap in a small patch of sky in Stripe 82. We want to quantify the benefits of combining SDSS spec-$z$ galaxies with the DES photometric galaxies in the Stripe 82 region. 

		We first note that if one wishes to combine DES and SDSS cosmological constraints correctly, computing the survey cross-correlations in Stripe82 is not optional, since DES and SDSS sample some of the same cosmological modes on the overlap patch. Our framework provides an easy way to compute those cross-correlations and the resulting combined cosmological constraints of the two surveys.

		However, because the overlap region is small relative to the full survey sizes, we do not expect direct cosmological parameter constraints from Stripe 82 to be significant relative to the full constraining power of the rest of DES and SDSS. Therefore any error from neglecting the cross-correlation should be small. Similarly, any improvement in direct cosmological parameter constraints from including the cross-correlation, such as that found for the LSST and DESI overlap, will be insignificant relative to the constraining power of the rest of the surveys.

		Nonetheless, we can hope that nuisance parameter constraints in Stripe 82 will translate outside of the patch to the rest of DES and improve cosmological constraints there. In particular, photo-$z$ errors are an important source of uncertainty for DES. Therefore, we would like to quantify the photo-$z$ constraining power of the overlap in Stripe 82 and compare it to constraints from a similarly sized disjoint patch of sky. Because we expect weak lensing to have little photo-$z$ constraining power, we can do this analysis considering galaxy clustering only.

		\begin{figure*}
		\includegraphics[width=\linewidth]{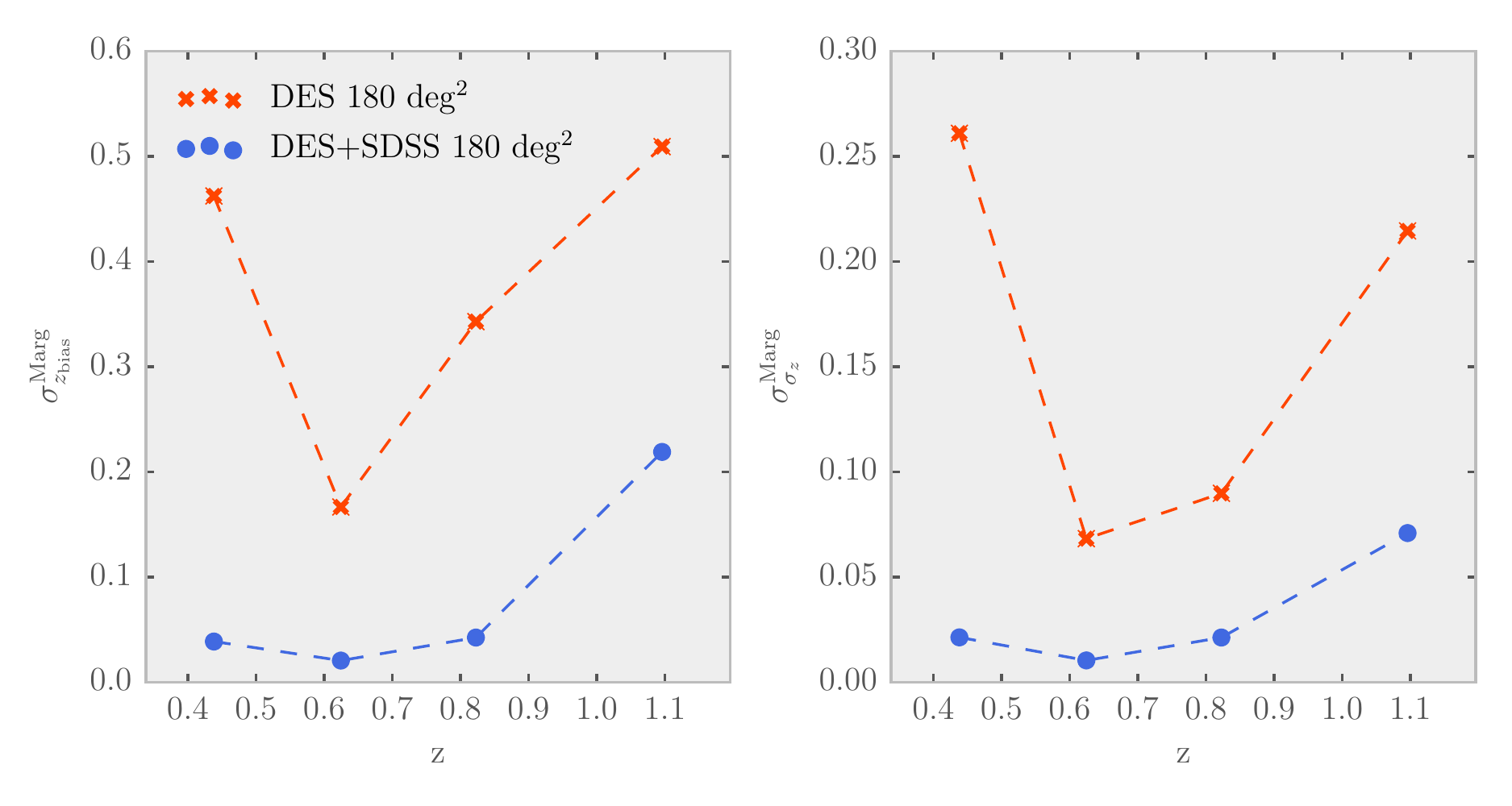}
		\caption{Errors on DES photo-$z$ parameters on a patch overlapping with SDSS (blue circles) and on an isolated patch (red crosses), with cosmological parameters fixed. We connect the discrete parameter constraints with dashed lines as a visual aid. Note that the scale is linear, and each panel has a different scale. Results are generally very similar to the LSST-DESI forecast, with same-sky improvement factors ranging from \numbers{{\sim }2} to \numbers{{\sim }12}. Just as in the LSST case, marginalizing over cosmological parameters degrades only the internal reconstruction points, not the cross-correlation points.}
		\label{fig:stripe82 photoz}
		\end{figure*}

		We see in Fig.~\ref{fig:stripe82 photoz} that a combined 2D$\times$3D analysis, including the field cross-correlation, drastically improves constraints on the redshift parameters of the photometric survey when compared to an analysis of disjoint patches of the same size, i.e. a DES self-calibrated photo-$z$ reconstruction. By cross-correlating the fields in our framework, marginalized errors on the photometric redshift bias parameters $z_\text{bias}$ and $\sigma_z$ are improved by factors ranging from \numbers{{\sim}2} at high-$z$, where there is little overlap between the DES and SDSS samples, to \numbers{{\sim}12} at low-$z$, where there is significant DES/SDSS overlap. Just as in the LSST-DESI case, we have fixed cosmological parameters so as to focus on photo-$z$ constraints derived from the cross-correlation with the 3D sample.

		\begin{figure}
		\includegraphics[width=\linewidth]{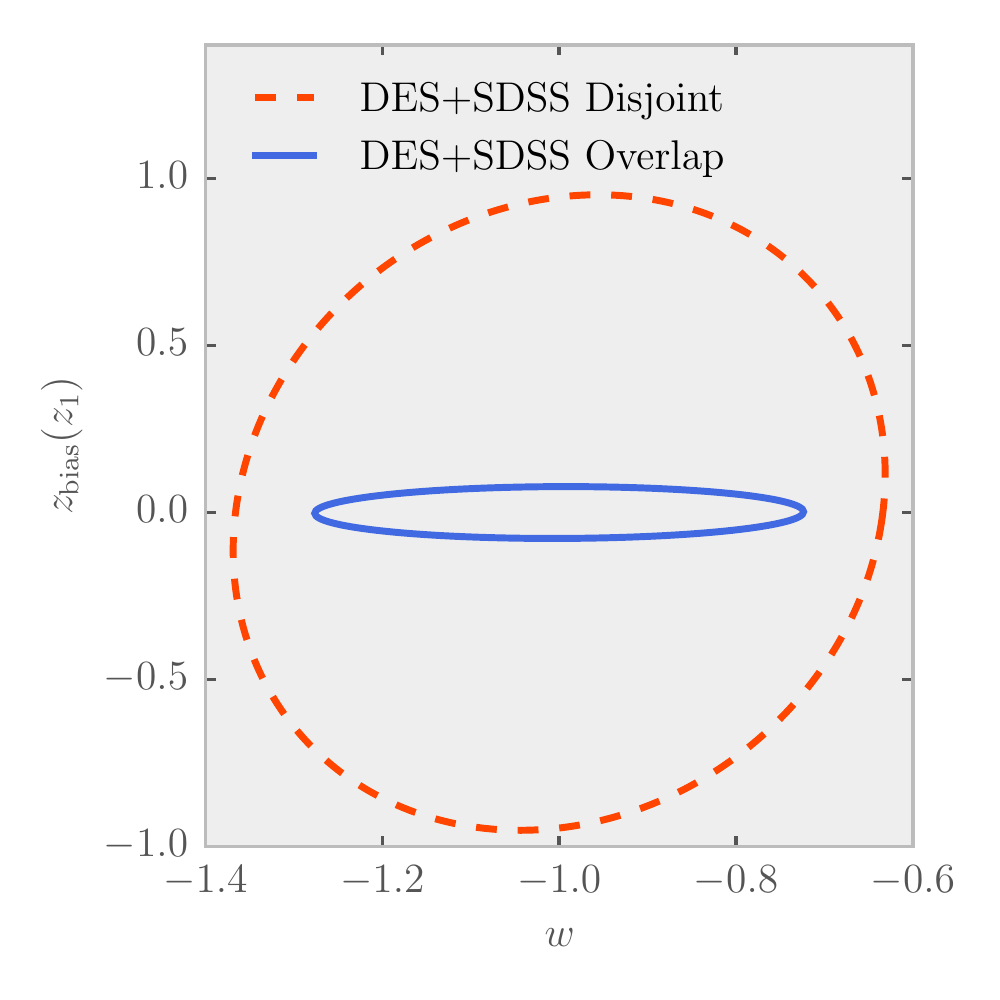}
		\caption{Simultaneous marginalized constraints from DES and SDSS on $w$ and the photo-$z$ bias parameter in the first bin, $z_{\text{bias}}(z_1)$. The smaller solid blue ellipse shows constraints from the DES and SDSS overlap in Stripe82, while the larger dashed red ellipse shows constraints on disjoint patches of the same size. Constraints on photo-$z$ parameters improve significantly when the cross-correlation is used, while constraints on cosmological parameters show more modest improvement. Combined constraint ellipses like this one are only possible in one-step combined analysis frameworks like the one presented here.}
		\label{fig:stripe82 zbias w0}
		\end{figure}

		\mathchardef\mhyphen="2D

		Our estimated constraints of \numbers{{\sim}0.03} on the photo-$z$ bias errors using our framework indicate that correlating SDSS spectroscopic galaxies with DES in Stripe 82 could be quite useful.
		We propose that such an analysis could profitably be done in the framework we have presented here, decomposing SDSS in sFB modes and DES in spherical harmonics. In this way, one can appropriately account for bias, cosmology, and photo-$z$ covariances while retaining all radial information in SDSS. Our framework allows the production of joint cosmology-photo-$z$ constraints, such as the joint covariance on constraints in the $w \mhyphen z_{\text{bias}}(z_1)$ plane illustrated in Fig.~\ref{fig:stripe82 zbias w0}, without any loss of information.

\section{Conclusion}
\label{sec:conclusion}

	We have presented a framework in which to jointly analyze overlapping 2D and 3D galaxy samples without loss of information. The framework exploits overlapping photometric and spectroscopic surveys to simultaneously constrain photometric redshift error parameters, galaxy bias parameters, and cosmological parameters. The framework is simple, theoretically well motivated, and data ready. Codes already exist to perform the decompositions we propose.

	As an example, we forecasted the benefits of the LSST-DESI overlap region and showed that the survey overlap will lead to improvements both in direct cosmological constraints and in photo-$z$ constraints. We accounted for bias evolution and photo-$z$ uncertainties, and we incorporated all available modes in both galaxy clustering and weak lensing. We did not model cosmological constraints from the disjoint survey regions, nor did we include CMB priors, covariance with overlapping CMB surveys, and RSDs, and thus we have not conducted a truly complete forecast of the benefits of an LSST/DESI overlap region. Rather, we showed that the sFB formalism is a fruitful framework in which such an analysis can be done.

	We also used the framework to show that the cross-correlation of SDSS-III and DES on Stripe 82 can place strong constraints on DES photo-$z$ parameters, and we emphasized the importance of a combined analysis framework when conducting this cross-correlation.

	While the framework we have presented here is ready for application on data, our work suggests several avenues of theoretical development. 

	In particular, the strong photo-$z$ constraints that we showed are enabled by survey overlap highlight that more work remains to be done on the best way to translate photo-$z$ constraints from the overlap patch to the rest of the photometric survey footprint. Such a calibration is essential to obtaining the strongest possible cosmological constraints with upcoming spectroscopic and photometric surveys.

	More broadly, the sFB decompositions we used here for spectroscopic surveys are applicable to a wide array of issues related to galaxy surveys. In particular, the separation of radial and angular basis functions mean the sFB decomposition might provide a useful framework to describe photometric redshift surveys themselves. Photometric surveys often contain subsets of objects with relatively well behaved photo-$z$ errors, such as, e.g., the redMaGiC galaxies in DES. Rather than binning these photometric galaxies into 2D slices, it might be possible to represent them directly in the sFB framework. This would require incorporating photo-$z$ errors into the sFB framework, which \cite{Kitching2011} have explored for shear fields. Extending this work to clustering fields has the potential to extract more radial information from photometric redshift surveys and therefore augment their ability to constrain late-universe physics.

\section*{Acknowledgements}

	We are grateful to Olivier Dor\'{e}, Alan Heavens, Donghui Jeong, Donnacha Kirk, Patrick McDonald, Brice M\'{e}nard, Jeffrey Newman, Marco Raveri, Hee-Jong Seo, An\v{z}e Slosar, and Masahiro Takada for helpful discussions and suggestions. We thank the developers of CosmicPy, which we used to compute the sFB window function. We are especially grateful to the referee, whose comments and suggestions were very useful.

	This work was partially supported by the Kavli Institute for Cosmological Physics at the University of Chicago through Grant No.~NSF PHY-1125897 and an endowment from the Kavli Foundation and its founder Fred Kavli. The work of S.D. is supported by the U.S. Department of Energy, including Grant No.~DE-FG02-95ER40896. This project used public archival data from the Dark Energy Survey (DES) and the Sloan Digital Sky Survey (SDSS).

\bibliographystyle{mnrasunsrt}
\bibliography{references}

\begin{thebibliography}{}
\makeatletter
\relax
\def\mn@urlcharsother{\let\do\@makeother \do\$\do\&\do\#\do\^\do\_\do\%\do\~}
\def\mn@doi{\begingroup\mn@urlcharsother \@ifnextchar [ {\mn@doi@}
  {\mn@doi@[]}}
\def\mn@doi@[#1]#2{\def\@tempa{#1}\ifx\@tempa\@empty \href
  {http://dx.doi.org/#2} {doi:#2}\else \href {http://dx.doi.org/#2} {#1}\fi
  \endgroup}
\def\mn@eprint#1#2{\mn@eprint@#1:#2::\@nil}
\def\mn@eprint@arXiv#1{\href {http://arxiv.org/abs/#1} {{\tt arXiv:#1}}}
\def\mn@eprint@dblp#1{\href {http://dblp.uni-trier.de/rec/bibtex/#1.xml}
  {dblp:#1}}
\def\mn@eprint@#1:#2:#3:#4\@nil{\def\@tempa {#1}\def\@tempb {#2}\def\@tempc
  {#3}\ifx \@tempc \@empty \let \@tempc \@tempb \let \@tempb \@tempa \fi \ifx
  \@tempb \@empty \def\@tempb {arXiv}\fi \@ifundefined
  {mn@eprint@\@tempb}{\@tempb:\@tempc}{\expandafter \expandafter \csname
  mn@eprint@\@tempb\endcsname \expandafter{\@tempc}}}

\bibitem[\protect\citeauthoryear{Weinberg}{Weinberg}{1989}]{Weinberg1989}
Weinberg S.,  1989, \mn@doi [Rev. Mod. Phys.] {10.1103/RevModPhys.61.1}, 61, 1

\bibitem[\protect\citeauthoryear{Feng}{Feng}{2010}]{Feng2010}
Feng J.~L.,  2010, \mn@doi [Annu. Rev. Astron. Astrophys.]
  {10.1146/annurev-astro-082708-101659}, 48, 495

\bibitem[\protect\citeauthoryear{Chang et~al.,}{Chang et~al.}{2016}]{Chang2016}
Chang C.,  et~al., 2016, \mn@doi [Mon. Not. R. Astron. Soc.]
  {10.1093/mnras/stw861}, 459, 3203

\bibitem[\protect\citeauthoryear{{Dark Energy Survey Collaboration}}{{Dark
  Energy Survey Collaboration}}{2005}]{TheDarkEnergySurveyCollaboration2005}
{Dark Energy Survey Collaboration} 2005, preprint (\mn@eprint {arXiv}
  {0510346})

\bibitem[\protect\citeauthoryear{Alam et~al.,}{Alam et~al.}{2015}]{Alam2015}
Alam S.,  et~al., 2015, \mn@doi [Astrophys. J. Suppl. Ser.]
  {10.1088/0067-0049/219/1/12}, 219, 12

\bibitem[\protect\citeauthoryear{{LSST Science Collaboration}}{{LSST Science
  Collaboration}}{2009}]{LSSTScienceCollaboration2009}
{LSST Science Collaboration} 2009, preprint (\mn@eprint {arXiv} {0912.0201})

\bibitem[\protect\citeauthoryear{{DESI Collaboration}}{{DESI
  Collaboration}}{2016}]{DESICollaboration2016}
{DESI Collaboration} 2016, preprint (\mn@eprint {arXiv} {1611.00036})

\bibitem[\protect\citeauthoryear{Newman et~al.,}{Newman
  et~al.}{2008}]{Newman2008}
Newman J.~A.,  et~al., 2008, \mn@doi [Astrophys. J.] {10.1086/589982}, 684, 88

\bibitem[\protect\citeauthoryear{McQuinn \& White}{McQuinn \&
  White}{2013}]{McQuinn2013}
McQuinn M.,  White M.,  2013, \mn@doi [Mon. Not. R. Astron. Soc.]
  {10.1093/mnras/stt914}, 433, 2857

\bibitem[\protect\citeauthoryear{M{\'{e}}nard, Scranton, Schmidt, Morrison,
  Jeong, Budavari  \& Rahman}{M{\'{e}}nard et~al.}{2013}]{Menard2013}
M{\'{e}}nard B.,  Scranton R.,  Schmidt S.,  Morrison C.,  Jeong D.,  Budavari
  T.,   Rahman M.,  2013, preprint (\mn@eprint {arXiv} {1303.4722})

\bibitem[\protect\citeauthoryear{Gazta{\~{n}}aga, Eriksen, Crocce, Castander,
  Fosalba, Marti, Miquel  \& Cabr{\'{e}}}{Gazta{\~{n}}aga
  et~al.}{2012}]{Gaztanaga2012}
Gazta{\~{n}}aga E.,  Eriksen M.,  Crocce M.,  Castander F.~J.,  Fosalba P.,
  Marti P.,  Miquel R.,   Cabr{\'{e}} A.,  2012, \mn@doi [Mon. Not. R. Astron.
  Soc.] {10.1111/j.1365-2966.2012.20613.x}, 422, 2904

\bibitem[\protect\citeauthoryear{Kirk, Lahav, Bridle, Jouvel, Abdalla  \&
  Frieman}{Kirk et~al.}{2013}]{Kirk2013}
Kirk D.,  Lahav O.,  Bridle S.,  Jouvel S.,  Abdalla F.~B.,   Frieman J.~A.,
  2013, preprint (\mn@eprint {arXiv} {1307.8062})

\bibitem[\protect\citeauthoryear{Eriksen \& Gaztanaga}{Eriksen \&
  Gaztanaga}{2015}]{Eriksen2015b}
Eriksen M.,  Gaztanaga E.,  2015, \mn@doi [Mon. Not. R. Astron. Soc.]
  {10.1093/mnras/stv1093}, 451, 1553

\bibitem[\protect\citeauthoryear{de Putter, Dor{\'{e}}  \& Takada}{de~Putter
  et~al.}{2013}]{DePutter2013}
de Putter R.,  Dor{\'{e}} O.,   Takada M.,  2013, preprint (\mn@eprint {arXiv}
  {1308.6070})

\bibitem[\protect\citeauthoryear{Font-Ribera, McDonald, Mostek, Reid, Seo  \&
  Slosar}{Font-Ribera et~al.}{2014}]{Font-Ribera2014}
Font-Ribera A.,  McDonald P.,  Mostek N.,  Reid B.~A.,  Seo H.-J.,   Slosar A.,
   2014, \mn@doi [J. Cosmol. Astropart. Phys.] {10.1088/1475-7516/2014/05/023},
  2014, 023

\bibitem[\protect\citeauthoryear{Fisher, Lahav, Hoffman, Lynden-Bell  \&
  Zaroubi}{Fisher et~al.}{1994}]{Fisher1994}
Fisher K.,  Lahav O.,  Hoffman Y.,  Lynden-Bell D.,   Zaroubi S.,  1994,
  \mn@doi [Mon. Not. R. Astron. Soc.] {10.1093/mnras/272.4.885}, 272, 885

\bibitem[\protect\citeauthoryear{Zaroubi, Hoffman, Fisher  \& Lahav}{Zaroubi
  et~al.}{1995}]{Zaroubi1995}
Zaroubi S.,  Hoffman Y.,  Fisher K.~B.,   Lahav O.,  1995, \mn@doi [Astrophys.
  J.] {10.1086/176070}, 449, 446

\bibitem[\protect\citeauthoryear{Heavens \& Taylor}{Heavens \&
  Taylor}{1995}]{Heavens1995}
Heavens A.~F.,  Taylor A.~N.,  1995, \mn@doi [Mon. Not. R. Astron. Soc.]
  {10.1093/mnras/275.2.483}, 275, 483

\bibitem[\protect\citeauthoryear{Shapiro, Crittenden  \& Percival}{Shapiro
  et~al.}{2012}]{Shapiro2011}
Shapiro C.,  Crittenden R.~G.,   Percival W.~J.,  2012, \mn@doi [Mon. Not. R.
  Astron. Soc.] {10.1111/j.1365-2966.2012.20785.x}, 422, 2341

\bibitem[\protect\citeauthoryear{Dai, Kamionkowski  \& Jeong}{Dai
  et~al.}{2012}]{Dai2012}
Dai L.,  Kamionkowski M.,   Jeong D.,  2012, \mn@doi [Phys. Rev. D]
  {10.1103/PhysRevD.86.125013}, 86, 125013

\bibitem[\protect\citeauthoryear{Rassat \& Refregier}{Rassat \&
  Refregier}{2012}]{Rassat2012}
Rassat A.,  Refregier A.,  2012, \mn@doi [Astron. Astrophys.]
  {10.1051/0004-6361/201118638}, 540, A115

\bibitem[\protect\citeauthoryear{Pratten \& Munshi}{Pratten \&
  Munshi}{2014}]{Pratten2014}
Pratten G.,  Munshi D.,  2014, \mn@doi [Mon. Not. R. Astron. Soc.]
  {10.1093/mnras/stu807}, 442, 759

\bibitem[\protect\citeauthoryear{Lanusse, Rassat  \& Starck}{Lanusse
  et~al.}{2015}]{Lanusse2015}
Lanusse F.,  Rassat A.,   Starck J.~L.,  2015, \mn@doi [Astron. Astrophys.]
  {10.1051/0004-6361/201424456}, 578, A10

\bibitem[\protect\citeauthoryear{Gorski, Hivon, Banday, Wandelt, Hansen,
  Reinecke  \& Bartelmann}{Gorski et~al.}{2005}]{Gorski2005}
Gorski K.~M.,  Hivon E.,  Banday A.~J.,  Wandelt B.~D.,  Hansen F.~K.,
  Reinecke M.,   Bartelmann M.,  2005, \mn@doi [Astrophys. J.]
  {10.1086/427976}, 622, 759

\bibitem[\protect\citeauthoryear{Hearin, Zentner  \& Ma}{Hearin
  et~al.}{2012}]{Hearin2012}
Hearin A.~P.,  Zentner A.~R.,   Ma Z.,  2012, \mn@doi [J. Cosmol. Astropart.
  Phys.] {10.1088/1475-7516/2012/04/034}, 2012, 034

\bibitem[\protect\citeauthoryear{Loverde \& Afshordi}{Loverde \&
  Afshordi}{2008}]{Loverde2008}
Loverde M.,  Afshordi N.,  2008, \mn@doi [Phys. Rev. D]
  {10.1103/PhysRevD.78.123506}, 78

\bibitem[\protect\citeauthoryear{Leistedt, Rassat, R{\'{e}}fr{\'{e}}gier  \&
  Starck}{Leistedt et~al.}{2012}]{Leistedt2012}
Leistedt B.,  Rassat a.,  R{\'{e}}fr{\'{e}}gier a.,   Starck J.-L.,  2012,
  \mn@doi [Astron. Astrophys.] {10.1051/0004-6361/201118463}, 540, A60

\bibitem[\protect\citeauthoryear{Yoo \& Desjacques}{Yoo \&
  Desjacques}{2013}]{Yoo2013}
Yoo J.,  Desjacques V.,  2013, \mn@doi [Phys. Rev. D]
  {10.1103/PhysRevD.88.023502}, 88, 023502

\bibitem[\protect\citeauthoryear{Takahashi, Sato, Nishimichi, Taruya  \&
  Oguri}{Takahashi et~al.}{2012}]{Takahashi2012}
Takahashi R.,  Sato M.,  Nishimichi T.,  Taruya A.,   Oguri M.,  2012, \mn@doi
  [Astrophys. J.] {10.1088/0004-637X/761/2/152}, 761, 152

\bibitem[\protect\citeauthoryear{Tegmark, Taylor  \& Heavens}{Tegmark
  et~al.}{1997}]{Tegmark1997}
Tegmark M.,  Taylor A.~N.,   Heavens A.~F.,  1997, \mn@doi [Astrophys. J.]
  {10.1086/303939}, 480, 22

\bibitem[\protect\citeauthoryear{Ade et~al.,}{Ade et~al.}{2016}]{Ade2016}
Ade P. A.~R.,  et~al., 2016, \mn@doi [Astron. Astrophys.]
  {10.1051/0004-6361/201525830}, 594, A13

\bibitem[\protect\citeauthoryear{Lewis, Challinor  \& Lasenby}{Lewis
  et~al.}{2000}]{Lewis2000}
Lewis A.,  Challinor A.,   Lasenby A.,  2000, \mn@doi [Astrophys. J.]
  {10.1086/309179}, 538, 473

\bibitem[\protect\citeauthoryear{Smith et~al.,}{Smith et~al.}{2003}]{Smith2003}
Smith R.~E.,  et~al., 2003, \mn@doi [Mon. Not. R. Astron. Soc.]
  {10.1046/j.1365-8711.2003.06503.x}, 341, 1311

\bibitem[\protect\citeauthoryear{Mead, Peacock, Heymans, Joudaki  \&
  Heavens}{Mead et~al.}{2015}]{Mead2015}
Mead A.~J.,  Peacock J.~A.,  Heymans C.,  Joudaki S.,   Heavens A.~F.,  2015,
  \mn@doi [Mon. Not. R. Astron. Soc.] {10.1093/mnras/stv2036}, 454, 1958

\bibitem[\protect\citeauthoryear{Ma, Hu  \& Huterer}{Ma et~al.}{2006}]{Ma2006}
Ma Z.,  Hu W.,   Huterer D.,  2006, \mn@doi [Astrophys. J.] {10.1086/497068},
  636, 21

\bibitem[\protect\citeauthoryear{de Putter, Dor{\'{e}}  \& Das}{de~Putter
  et~al.}{2014}]{DePutter2014}
de Putter R.,  Dor{\'{e}} O.,   Das S.,  2014, \mn@doi [Astrophys. J.]
  {10.1088/0004-637X/780/2/185}, 780, 185

\bibitem[\protect\citeauthoryear{Bonnett et~al.,}{Bonnett
  et~al.}{2016}]{Bonnett2016}
Bonnett C.,  et~al., 2016, \mn@doi [Phys. Rev. D] {10.1103/PhysRevD.94.042005},
  94, 042005

\bibitem[\protect\citeauthoryear{Cacciato, Lahav, van~den Bosch, Hoekstra  \&
  Dekel}{Cacciato et~al.}{2012}]{Cacciato2012}
Cacciato M.,  Lahav O.,  van~den Bosch F.~C.,  Hoekstra H.,   Dekel A.,  2012,
  \mn@doi [Mon. Not. R. Astron. Soc.] {10.1111/j.1365-2966.2012.21762.x}, 426,
  566

\bibitem[\protect\citeauthoryear{Jouvel, Abdalla, Kirk, Lahav, Lin, Annis, Kron
   \& Frieman}{Jouvel et~al.}{2014}]{Jouvel2014}
Jouvel S.,  Abdalla F.~B.,  Kirk D.,  Lahav O.,  Lin H.,  Annis J.,  Kron R.,
  Frieman J.~A.,  2014, \mn@doi [Mon. Not. R. Astron. Soc.]
  {10.1093/mnras/stt2371}, 438, 2218

\bibitem[\protect\citeauthoryear{Kitching, Heavens  \& Miller}{Kitching
  et~al.}{2011}]{Kitching2011}
Kitching T.~D.,  Heavens A.~F.,   Miller L.,  2011, \mn@doi [Mon. Not. R.
  Astron. Soc.] {10.1111/j.1365-2966.2011.18369.x}, 413, 2923

\bibitem[\protect\citeauthoryear{Heavens, Kitching  \& Taylor}{Heavens
  et~al.}{2006}]{Heavens2006}
Heavens A.~F.,  Kitching T.~D.,   Taylor A.~N.,  2006, \mn@doi [Mon. Not. R.
  Astron. Soc.] {10.1111/j.1365-2966.2006.11006.x}, 373, 105

\makeatother
\end{thebibliography}

\appendix 

\section{Fiducial Cosmology}
	\label{app: fiducial}
	A fiducial redshift-distance relation $r_{\text{fid}}(z) \equiv \tilde{r}$ is required to convert a spectroscopic survey's redshift data into the comoving galaxy overdensity. In general, this redshift-distance relation will not be the cosmologically correct one. \cite{Heavens2006} emphasized that this error will degrade cosmological constraints. Any $P(k)$ style analysis will suffer from this degradation. An analysis that remains completely in redshift space will preserve all information, but such analyses are hampered by the large number of redshift bins needed to fully conserve the radial information in the 3D field.

	With an assumed redshift-distance relation, the data in hand from a spectroscopic redshift survey is not the true 3D overdensity $\beta(r \hat{n})$, but rather an estimated one  $\hat{\beta}(\tilde{r} \hat{n})$, related to the true overdensity by $\hat{\beta}(\tilde{r} \hat{n})$ = $\beta(r \hat{n})$. Expanding the estimated field $\hat{\beta}(\tilde{r} \hat{n})$ in the sFB basis, we have:
	\begin{equation}
		\beta(r \hat{n}) = \hat{\beta}(\tilde{r} \hat{n}) = \sqrt{\frac{2}{\pi}} \sum_{\ell,m} \int_0^\infty q \ \diff{q} \ \beta_{\ell m} (q) Y_{\ell m} (\hat{n}) j_\ell (q\tilde{r}).
	\end{equation}
	Note the presence of the fiducial distance in the Bessel function argument. We again solve for the expansion coefficients, 
	\begin{align*}
		\beta_{\ell m}&(q) = \sqrt{\frac{2}{\pi}} \int_0^\infty \tilde{r}^2 \diff{\tilde{r}} \int \diff{\Omega} \bigg[ \\ &q Y^*_{\ell m} (\hat{n}) j_\ell (q \tilde{r}) \phi_\beta(r) b_\beta(r) \delta(r\hat{n}) \bigg],
		\numberthis
	\end{align*}
	and take the Fourier transform,
	\begin{equation}
		\beta_{\ell m}(q) = \sqrt{\frac{2}{\pi}} \int \frac{ \diffcubed{\vec{k}}}{(2 \pi)^3} \tilde\delta(\vec{k}) (4 \pi) i^\ell Y^*_{\ell m} (\hat{k}) W^\beta_{\ell}(k,q).
	\end{equation}
	We find that the effect of the fiducial cosmology can be entirely absorbed into the window $W^\beta_{\ell}(k,q)$:
	\begin{equation}
		W^\beta_{\ell}(k,q) \equiv q \int \tilde{r}^2 \diff{\tilde{r}} j_\ell (q\tilde{r}) j_\ell (kr) D(r, k) b_\beta(r) \phi_\beta(r).
	\end{equation}
	The specific way in which the fiducial distance relation appears here is different than in \cite{Lanusse2015}. Though a priori this difference could have some impact on our constraints, in practice we find it only affects the constraint on $w$, and that by at most a few percent.

\label{lastpage}
\end{document}